\documentclass[aps,prl,showpacs,twocolumn,superscriptaddress]{revtex4}
\usepackage{bm,color}
\usepackage{graphicx}
\usepackage{amsmath}

\begin{document}
\title {Theoretically predicted picosecond optical switching of spin 
chirality in multiferroics}

\author{Masahito Mochizuki}
\affiliation{Department of Applied Physics, University of Tokyo, 
Tokyo 113-8656, Japan}
\affiliation{Multiferroics Project, ERATO, Japan Science and Technology 
Agency (JST), Tokyo 113-8656, Japan}

\author{Naoto Nagaosa}
\affiliation{Department of Applied Physics, University of Tokyo,
Tokyo 113-8656, Japan}
\affiliation{Cross-Correlated Materials Research Group (CMRG)
and Correlated Electron Research Group (CERG),
RIKEN-ASI, Saitama 351-0198, Japan}

\begin{abstract}
We show theoretically with an accurate spin Hamiltonian describing the multiferroic Mn perovskites that the application of the picosecond optical pulse with a terahertz frequency can switch the spin chirality through intensely exciting the electromagnons. There are four states with different spin chiralities, i.e. clockwise and counterclockwise $ab$/$bc$-plane spin spirals, and by tuning the strength, shape and length of the pulse, the switching among these states can be controlled at will. Dynamical pattern formation during the switching is also discussed.
\end{abstract}
\maketitle
\begin{figure}[tdp]
\includegraphics[scale=1.0]{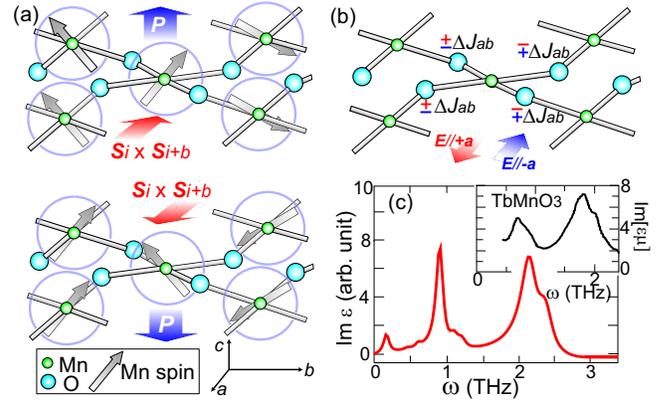}
\caption{(Color online) (a) Spin configuration, spin chirality 
$\bm S_i \times \bm S_{i+b}$ and 
ferroelectric polarization $\bm P$ in the 
clockwise $bc$-plane spin spiral (upper panel) and those in the 
counterclockwise one (lower panel). 
(b) Modulation of the in-plane ferromagnetic exchanges under 
$\bm E$$\parallel$$\pm \bm a$.
Upper (lower) signs in front of $\Delta J_{ab}$ ($>$0) correspond to 
the modulations under $\bm E$$\parallel$$+\bm a$ 
($\bm E$$\parallel$$-\bm a$).
(c) Calculated electromagnon spectrum. Inset shows the experimental 
spectrum for TbMnO$_3$~\cite{Takahashi08}.}
\label{Fig01}
\end{figure}

Chirality, i.e., the right- and left-handedness of structure, is one of the 
fundamental concepts penetrating through the whole of science.
In solids, electron spins sometimes form a chiral order, 
which offers an opportunity to manipulate the chirality by external 
parameters.
This issue is of vital importance in spintronics, which aims at the 
electric control of spins by an electric current or electric field ($\bm E$).
Multiferroics provides us an ideal system for this purpose, in which
the spin chirality is directly related to the electric 
polarization~\cite{MFReview,Kimura03a}.
In the spin-current model~\cite{Katsura05,Mostovoy06}, two mutually 
canted spins $\bm S_i$ and $\bm S_j$ generate the polarization 
$\bm p_{ij}$ as
\begin{equation}
\bm p_{ij} \propto \bm e_{ij} \times (\bm S_i \times \bm S_j),
\label{eq:PScrsS}
\end{equation}
with $\bm e_{ij}$ being the vector connecting $i$th and $j$th sites.
Here the vector product $\bm S_i \times \bm S_j$ is the spin chirality, 
which characterizes a direction of the spin rotation. 
In multiferroic Mn perovskites such as TbMnO$_3$ and DyMnO$_3$,
the Mn spins are rotating within the $bc$ plane to form a cycloid 
($bc$-plane spiral) propagating along the $+\bm b$ direction 
($Pbnm$ setting) as shown in 
Fig.~\ref{Fig01}(a)~\cite{Kenzelmann05,Yamasaki07}, and the clockwise 
(CW) one with ($\bm S_i \times \bm S_j$)$\parallel$$-\bm a$
induces $\bm P$$\parallel$$+\bm c$, while the counterclockwise 
(CCW) one with ($\bm S_i \times \bm S_j$)$\parallel$$+\bm a$
induces $\bm P$$\parallel$$-\bm c$. 

In this Letter, we theoretically propose
a picosecond optical switching of the spin chirality in 
TbMnO$_3$ as a representative material. However, the mechanism is rather
general and would be ubiquitous in spiral magnets.
It is found that the oscillating $\bm E$ component
of the optical pulse activates collective rotations of the spin-spiral 
planes via magnetoelectric (ME) coupling, 
and their inertial motions result in the chirality reversal or flop. 
By tuning the strength, shape and length of the pulse, 
the spin chirality is shown to be controlled at will.

When the electric polarization is driven by the spin order, it is 
naturally expected that the collective excitation of spins (magnon) has 
an infrared activity~\cite{Pimenov06,Katsura07}. 
Indeed, strong optical absorptions were experimentally observed in 
$R$MnO$_3$ ($R$=Tb, Dy, Eu$_{1-x}$Y$_x$, etc) at THz frequencies, 
and they were ascribed to magnons activated by the $\bm E$ component 
of the light, i.e., electromagnons~\cite{Pimenov06,Pimenov06b,Pimenov08,
Kida08,Kida08b,Takahashi08,Kida09}. 
In the early stage, the corresponding magnon modes were interpreted 
as rotation of the spin-spiral plane with oscillating
$\bm p_{ij}$ in Eq.~(1)~\cite{Katsura07}. However, this 
interpretation contradicts the experimental observation about 
the selection rule in terms of the light 
polarization~\cite{Kida08,Kida08b}. Afterwards, it turned out that the
polarization $\bm p_{ij}$ due to the
 conventional magnetostriction, i.e., 
\begin{equation}
\bm p_{ij} =\bm \pi_{ij} (\bm S_i \cdot \bm S_j)
\label{eq:PSdotS}
\end{equation}
is relevant to the infrared-activity of magnons in 
$R$MnO$_3$~\cite{Aguilar09,MiyaharaCD,Mochizuki10}. Here the vector 
$\bm \pi_{ij}$ is nonzero because of the orthorhombic lattice 
distortion without inversion symmetry at the center of the Mn-O-Mn bond. 
The puzzling electromagnon optical spectrum with two specific peaks 
was successfully explained by this mechanism~\cite{Mochizuki10}. 

Under this circumstance, the photo-induced phenomena become a challenging 
issue. Since $\bm p_{ij}$ in Eq.~(2) does not require the spin-orbit 
interaction, its magnitude is much larger than that of $\bm p_{ij}$ in 
Eq.~(1), which enables the intense and fast optical excitation 
of magnons. This offers a unique opportunity to study the nonlinear
dynamics of the spin system.
In addition, the light can locally activate or modify the spin structure 
with a squeezed light spot in contrast to the magnetic field.

We are now ready to attack such phenomena in $R$MnO$_3$ 
theoretically ahead of experiments for the following reasons. 
First, we know that the 
optical pulse activates mostly the spins only via the 
ME coupling, which allows us to neglect 
electronic excitations at much higher energies ($>$1.5 eV). 
Second, we have an accurate spin Hamiltonian, which describes 
competitions among various phases in 
$R$MnO$_3$~\cite{Mochizuki09}, so that the optical 
switchings among them and dynamics after the light irradiation 
can be simulated in a reliable way. 

We employ a classical Heisenberg model on a cubic lattice, which 
contains not only the frustrating spin exchanges ($\mathcal{H}_{\rm ex}$)
but also the single-ion spin anisotropy ($\mathcal{H}_{\rm sia}^D$ and 
$\mathcal{H}_{\rm sia}^E$), Dzyaloshinskii-Moriya (DM) interaction 
($\mathcal{H}_{\rm DM}$) and biquadratic interaction 
($\mathcal{H}_{\rm biq}$) as,
\begin{eqnarray}
\mathcal{H}&=&\mathcal{H}_{\rm ex}
+\mathcal{H}_{\rm sia}^D+\mathcal{H}_{\rm sia}^E
+\mathcal{H}_{\rm DM}+\mathcal{H}_{\rm biq}  \nonumber \\
&=&\sum_{<i,j>} J_{ij} \bm S_i \cdot \bm S_j
+D \sum_{i} S_{\zeta i}^2  \nonumber \\
&+&E\sum_{i}(-1)^{i_x+i_y}(S_{\xi i}^2-S_{\eta i}^2) \nonumber \\
&+&\sum_{<i,j>}\bm d_{ij}\cdot(\bm S_i \times \bm S_j)
-B_{\rm biq}\sum_{<i,j>}^{ab}(\bm S_i \cdot \bm S_j)^2.
\end{eqnarray}
The frustration between ferromagnetic (FM) exchange
$J_{ab}$ and antiferromagnetic (AFM) exchange $J_b$ results in the 
in-plane spiral spin orders, while the interplane AFM exchange $J_c$ 
causes their staggered stacking. The DM vectors 
$\bm d_{i,j}$ are expressed using five DM parameters, 
$\alpha_{ab}$, $\beta_{ab}$, $\gamma_{ab}$, $\alpha_c$, and $\beta_c$, 
as given in Ref.~\cite{Solovyev96}.
Crucial roles of the biquadratic interaction in $R$MnO$_3$ were uncovered 
in recent theoretical studies~\cite{Kaplan09,Mochizuki10}. 
For more detail of the model, see Ref.~\cite{Mochizuki10}.
We adopt the following parameters: 
$J_{ab}$=$-$0.74, $J_b$=0.64, $J_c$=1.0, ($\alpha_{ab}$, $\beta_{ab}$, 
$\gamma_{ab}$)=(0.1, 0.1, 0.14), ($\alpha_c$, $\beta_c$)=(0.48, 0.1), 
$D$=0.2, $E$=0.25, and $B_{\rm biq}$=0.025, where the energy unit is meV. 
This parameter set gives the $bc$-plane spin spiral with 
a wave number $q_b$=0.3$\pi$ at low temperatures, which resembles the 
spin structure in TbMnO$_3$ ($q_b$=0.29$\pi$)~\cite{Kenzelmann05,Yamasaki07}.

We trace dynamics of the Mn spins by numerically solving the 
Landau-Lifshitz-Gilbert equation using the fourth-order Runge-Kutta 
method. We derive an effective magnetic field 
$\bm H^{\rm eff}_i$ acting on the spin $\bm S_i$ from the Hamiltonian 
$\mathcal{H}$ as 
$\bm H^{\rm eff}_i = - \partial \mathcal{H} / \partial \bm S_i$. 
Considering the observed reduced Mn moment~\cite{Arima06}, we set the 
norm of the spin vector $|\bm S_i|$=1.4. The system used for calculations 
is 40$\times$40$\times$6 in size with the periodic boundary condition.
For the ME coupling, we consider $-\bm E \cdot \bm p_{ij}$ 
with $\bm p_{ij}$ given in Eq.~(2)~\cite{Aguilar09,MiyaharaCD,Mochizuki10}.
This coupling effectively modulates the in-plane FM 
exchanges from $J_{ab} \bm S_i \cdot \bm S_j$ to 
$(J_{ab}-\bm E \cdot \bm \pi_{ij}) \bm S_i \cdot \bm S_j$. 
Consequently, the applied $\bm E$$\parallel$$\pm \bm a$ modulates 
the spin exchanges as shown in Fig.~\ref{Fig01}(b). 
Here $|\pi_{ij}^a|$ is calculated to be 3.5$\times$10$^{-26}$ $\mu Cm$ 
from the lattice parameters~\cite{Alonso00} and the observed ferroelectric 
polarization $P$($\sim$5000$\mu C/m^2$) for $R$MnO$_3$ with an 
up-up-down-down spin order~\cite{Ishiwata10}. 
This means that $E_a$=1 MV/cm induces the modulation 
$|\Delta J_{ab}|$=$|E_a \pi_{ij}^a|$=0.022 meV.

The infrared-absorption spectrum is calculated as the response to 
a weak $\delta$-function pulse.
(For technical detail, see Ref.~\cite{Mochizuki10}). 
The calculated spectrum is displayed in Fig.~\ref{Fig01}(c), which has 
two peaks at $\omega$=0.94 THz and 2.1 THz, and reproduces well the 
experimental spectrum of TbMnO$_3$. The Gilbert-damping coefficient 
$\alpha_{\rm G}$ is chosen to be 0.1 so as to reproduce the observed 
peak width, which guarantees the under-damped 
spin oscillations.

Now we theoretically demonstrate switching of the spin chirality by 
the optical pulse. There are two kinds of 
$bc$-plane spirals with different spin chiralities, i.e., CW and 
CCW ones. 
Their spin chiralities $\bm C$ point in the $-\bm a$ and $+\bm a$ 
directions ($\bm C$$\parallel$$-\bm a$ and $\bm C$$\parallel$$+\bm a$) 
so that they are referred to as $bc_{-}$ and $bc_{+}$, respectively. 
Here the chirality $\bm C$ is defined as a sum of the local contributions 
$\bm C_{i,i+\hat x}=\bm S_i \times \bm S_{i+\hat x}$ and 
$\bm C_{i,i+\hat y}=\bm S_i \times \bm S_{i+\hat y}$ as
$\bm C=\frac{1}{2N} \sum_{i} (\bm C_{i,i+\hat x} + \bm C_{i,i+\hat y})/S^2$.
Note that $\bm C_{i,i+\hat z}$ is zero because of the AFM 
stacking in $z$-direction.
The CW and CCW $ab$-plane spirals are also possible 
although they are slightly higher in energy than the 
$bc$-plane ones without external fields. 
They have $\bm C$$\parallel$$-\bm c$ and $\bm C$$\parallel$$+\bm c$, 
and thus are referred to as $ab_{-}$ and $ab_{+}$, respectively.

\begin{figure}[tdp]
\includegraphics[scale=1.0]{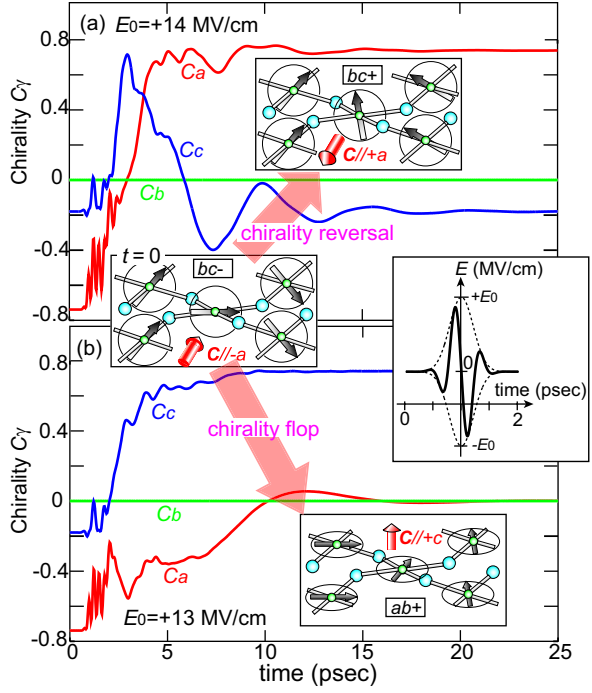}
\caption{Time evolutions of the spin chirality 
$\bm C$=($C_a$, $C_b$, $C_c$) after application of the pulse 
with (a) $E_0$=+14 MV/cm and (b) $E_0$=+13 MV/cm. 
Fig.2(a) shows the chirality-reversal from $bc_{-}$ ($C_a$$<$0) 
to $bc_{+}$ ($C_a$$>$0), while Fig.2(b) shows the chirality flop 
from $bc_{-}$ ($\bm C$$\parallel$$-\bm a$) to $ab_{+}$ 
($\bm C$$\parallel$$+\bm c$). 
Insets show spin states before or after applying the pulse, and time 
profile of the applied pulse $E_a(t)$.}
\label{Fig02}
\end{figure}
Starting with $bc_{-}$ with $\bm C$$\parallel$$-\bm a$, we apply an 
intense pulse of $\bm E$=$[E_a(t), 0, 0]$ along the $a$ axis, where
\begin{equation}
E_a(t)=-E_0 \sin \omega t \exp[-\frac{(t-t_0)^2}{2\sigma^2}],
\end{equation}
with $t_0$=1 psec. Here the frequency $\omega$ is 
fixed at 2.1 THz, which corresponds to the higher-energy 
electromagnon peak, while the full width of the half maximum for the 
Gaussian envelope, $2\sqrt{2 \ln2}\sigma$, is taken to be 0.5 
psec [see inset of Fig.~\ref{Fig02}]. 
In Fig.~\ref{Fig02}, we display calculated time evolutions of the 
$a$-, $b$- and $c$-axis components of $\bm C$. 
When $E_0$=+14 MV/cm, we find a reversal of $\bm C$ from $bc_{-}$ 
($C_a$$<$0) to $bc_{+}$ ($C_a$$>$0) as shown in Fig.~\ref{Fig02}(a). 
This reversal takes place via $ab_{+}$ with $C_c$$>$0 and $C_a$$\sim$0. 
In addition, as shown in Fig.~\ref{Fig02}(b), when we apply a slightly 
weaker pulse of $E_0$=+13 MV/cm, a chirality flop occurs from $bc_{-}$ 
($\bm C$$\parallel$$-\bm a$) to $ab_{+}$ ($\bm C$$\parallel$$+\bm c$). 
These switchings occur very fast, typically within 5-7 psec.

\begin{figure}[tdp]
\includegraphics[scale=1.0]{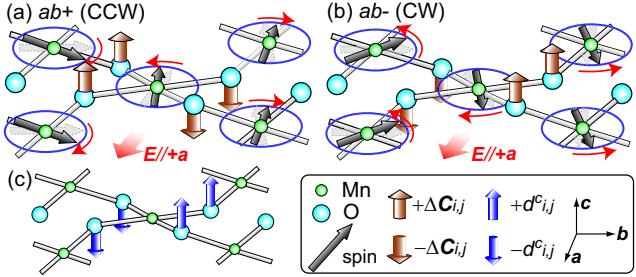}
\caption{Modulations of spin turn angles 
(thin red arrows) and the local spin chiralities 
$\pm \Delta \bm C_{i,j}$ under $\bm E$$\parallel$$+\bm a$ for 
(a) $ab_{+}$ and (b) $ab_{-}$ states.
(c) Arrangement of the $c$-axis components of DM vectors $\pm d_{i,j}^c$.}
\label{Fig03}
\end{figure}
To understand these phenomena, we first consider energies of the four 
chirality states, i.e., $ab_{\pm}$ and $bc_{\pm}$. 
When $\bm E$=0, the $ab_{+}$ and $ab_{-}$ are degenerate, and are higher 
in energy than the ground-state $bc$-plane spirals. 
Application of $\bm E$$\parallel$$\pm \bm a$ lifts this degeneracy, and 
the $ab_{+}$ ($ab_{-}$) becomes the lowest in energy under the 
strong $\bm E$$\parallel$$+\bm a$ ($\bm E$$\parallel$$-\bm a$).
This can be understood as follows. As shown in Figs.~\ref{Fig03}(a) and 
~\ref{Fig03}(b), the modified in-plane FM exchanges,
$J_{ab} \pm \Delta J_{ab}$, under $\bm E$$\parallel$$\pm \bm a$ cause
changes in the spin turn angles (thin red arrows) and hence staggered 
modulations of the local spin chiralities as 
$\bm C_{i,j} \pm \Delta \bm C_{i,j}$ (thick brown arrows).
Under $\bm E$$\parallel$$+\bm a$, the modulations 
$\pm \Delta \bm C_{i,j}$ in the $ab_{+}$ ($ab_{-}$) are always
antiparallel (parallel) to the staggered $c$-axis components of 
the DM vectors $\pm d_{i,j}^c$ [blue arrows in Fig.~\ref{Fig03}(c)]. 
Since the DM coupling favors the antiparallel configurations 
of $\pm d_{i,j}^c$ and $\pm \bm C_{i,j}$,
the energy decreases (increases) in the $ab_{+}$ ($ab_{-}$).
Note that this mechanism is distinct from the direct coupling 
between $\bm E$ and $\bm p_{ij}$ given in Eq.~(1), which is much 
weaker in the THz-frequency regime. 
On the other hand, the energies of $bc_{\pm}$ are not affected by 
$\bm E$ because the $a$-axis components of the DM vectors are 
alternately stacked, by which the DM energy always cancels out.

\begin{figure}[tdp]
\includegraphics[scale=1.0]{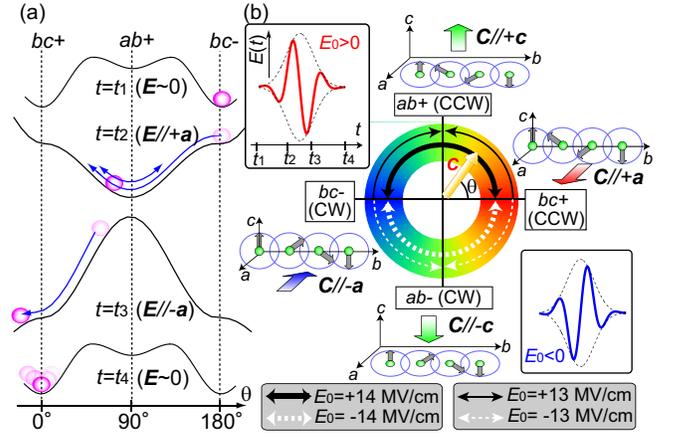}
\caption{
(a) Schematic figure for time evolution of the potential structure 
in the $\theta$ space during the chirality-reversal 
process for $E_0$=+14 MV/cm. Here $\theta$ is the angle between 
spin chirality $\bm C$ and the $a$ axis.
For a time profile of the applied pulse 
with $E_0$$>$0, see inset of (b).
(b) Relationships between the chirality-switching processes and the 
sign of $E_0$.}
\label{Fig04}
\end{figure}
In Fig.~\ref{Fig04}(a), we illustrate schematic time evolution of 
the potential as a function of $\theta$ during the chirality-reversal 
process. For time profile of the applied pulse with $E_0$$>$0, see 
the inset of Fig.~\ref{Fig04}(b). Here $\theta$ is the angle between 
the chirality $\bm C$ and the $a$ axis.
At $t$=$t_1$, the system is located in the minimum at 
$\theta$=180$^{\circ}$ ($bc_{-}$). When $E_a(t)$$>$0 as at $t$=$t_2$, 
$\theta$=90$^{\circ}$ ($ab_{+}$) becomes a new energy minimum, so that 
the chirality $\bm C$ starts rotating or the angle $\theta$ starts 
decreasing towards this minimum. 
Importantly the chirality does not stop its rotation at 
$\theta$=90$^{\circ}$ immediately, but passes through that minimum or 
oscillates around it because of the inertial force. 
The DM interaction and the single-ion anisotropy originating from 
the spin-orbit interaction make the rotation of spin-spiral plane 
massive, resulting in its inertial motion.
Then the $E_a(t)$ becomes negative as at $t$=$t_3$, which makes 
$ab_{+}$ ($\theta$=90$^{\circ}$) the highest in energy. 
Consequently the system starts falling into the minimum at 
$\theta$=0$^{\circ}$ ($bc_{+}$). 
At last ($t$=$t_4$), the chirality reversal is completed being 
settled in $bc_{+}$ ($\theta$=0$^{\circ}$).

On the other hand, the chirality flop for $E_0$=+13 MV/cm is a 
rather subtle process. If the system is within the domain of 
metastability of $\theta$=90$^{\circ}$ when $E_a(t)$ becomes almost 
zero at $t$=$t_4$, the system can be trapped in the local minimum 
of $ab_{+}$. Since the $ab_{\pm}$ are metastable, the system 
should decay into the ground-state $bc_{+}$ or $bc_{-}$ eventually 
due to thermal fluctuations. 

\begin{figure}[tdp]
\includegraphics[scale=1.0]{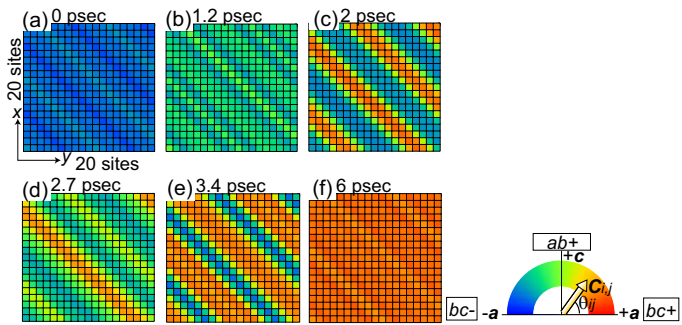}
\caption{(a)-(f) Color maps of calculated angles $\theta_{ij}$ between 
the local spin chiralities $\bm C_{i,j}$ and the $a$ axis (see inset), 
which show the real-time dynamics of $\bm C_{i,j}$. 
}
\label{Fig05}
\end{figure}
This ultrafast chirality switching is distinct from the conventional 
polarization reversal, and is accompanied by the dynamical spatial-pattern 
formation, i.e., dynamical stripes of chirality domains, through 
the nonlinear photoexcitation of electromagnons.
In Figs.~\ref{Fig05}(a)-(f), we show the calculated real-time dynamics of 
the local chiralities $\bm C_{i,j}$ on the Mn-O plane for the 
chirality-reversal process. Starting from $bc_{-}$ (blue) at $t$=0, 
the optical pulse generates $ab_{+}$ (green) domains as at $t$=1.2 psec. 
Subsequently stripes of $bc_{+}$ (red) and $bc_{-}$ (blue) domains emerge 
after the pulse ends at $t$=2 psec, and the chiralities $\bm C_{i,j}$ 
oscillate in each domain between $\theta_{ij}$=0$^{\circ}$ and 
$\theta_{ij}$=180$^{\circ}$ as seen in 2$<$$t$(psec)$<$3.4. 
Among the spirally ordered spins, those directing (nearly) along the 
propagation vector cannot flip and become nodes of the spin oscillations 
to form the chirality domains.
At last ($t$=6 psec) the system gradually 
gets settled in the $bc_{+}$ state.

Finally we discuss conditions for the chirality switching. 
First, the switching occurs only at a frequency of the higher-energy 
electromagnon resonance ($\omega$$\sim$2.1 THz in the present case), 
and does not occur at the lower-energy peak ($\omega$$\sim$1 THz).
Second, we need a rather large peak height of the pulse, 
$|E_0|$$\gtrsim$10 MV/cm, in the present simulation. 
Experimentally maximum peak height exceeding 100 MV/cm is available 
for 10-72 THz~\cite{Sell08}, but below 3 THz, it reaches only 
$\sim$1 MV/cm at present. 
We expect that the optical pulse with $|E_0|$$\gtrsim$10 MV/cm at 
$\sim$2 THz will be realized in the near future. 
Besides, the threshold value of $|E_0|$ can be reduced if we properly 
choose the target materials: For example, solid solutions 
Tb$_{1-x}$Gd$_x$MnO$_3$ locating near the boundary between the 
$ab$- and $bc$-plane spiral phases~\cite{Goto05} are promising candidates. 
Importantly there are optimal ranges of the electric strength $|E_0|$, 
and a larger $|E_0|$ cannot necessarily induce the switching. 
This can be understood as follows. To achieve the chirality reversal, 
for instance, from $\theta$=180$^{\circ}$ ($bc_{-}$) to 
$\theta$=0$^{\circ}$ ($bc_{+}$), the chirality vector oscillating around 
the energy minimum at $\theta$=90$^{\circ}$ ($ab_{+}$) should be 
in the range 0$^{\circ}$$<$$\theta$$<$90$^{\circ}$ when $\bm E$ 
is reversed from $E_a(t)$$>$0 to $E_a(t)$$<$0 in order to fall into the 
another minimum at $\theta$=0$^{\circ}$ ($bc_{+}$) instead of 
$\theta$=180$^{\circ}$ ($bc_{-}$). 
This means that we need to adjust depth of the energy minimum of 
$ab_{+}$ at $t$=$t_2$ by tuning the strength of $E_0$ in order to 
synchronize the timing between the chirality oscillation and the 
reversal of $\bm E$. 
Therefore, the switching processes show highly nonlinear behaviors 
with respect to strength and shape of the pulse. 
In addition, if we adopt a negative $E_0$ in Eq.~(4), the 
lowest-lying state at $t$=$t_2$ becomes $ab_{-}$ with 
$\theta$=270$^{\circ}$. 
Then the chirality reversal occurs via $ab_{-}$ instead of $ab_{+}$. 
The chirality flops to $ab_{-}$ from $bc_{\pm}$ become also possible 
for a slightly weaker $|E_0|$. 
Relationships between the switching processes and the sign of $E_0$ 
are summarized in Fig.~\ref{Fig04}(b).

To summarize, we have theoretically studied the 
ultrafast optical switching of spin chirality by exciting the 
electromagnons in the multiferroic Mn perovskite. 
We have revealed that the oscillating $\bm E$ component of the 
light activates the collective rotations of the spin-spiral planes with 
a THz frequency via the ME coupling, and their inertial 
motions result in chirality reversal or flop. 
By tuning strength, shape and length of the pulse, the spin chirality 
is shown to be controlled at will. 

The authors are grateful to N. Kida, Y. Tokura, N. Furukawa, 
R. Shimano and I. Kezmarki for discussions. 
This work was supported by Grant-in-Aid
(Grants No. 22740214, No. 21244053, No. 17105002, No. 19048015, 
and No. 19048008) and 
G-COE Program ``Physical Sciences Frontier" from MEXT Japan, 
and Funding Program for World-Leading Innovative R$\&$D on Science and 
Technology (FIRST Program) from JSPS.


\end{document}